\documentstyle [preprint,aps,epsf,eqsecnum,floats]{revtex}

\begin{document}

\begin{titlepage}
\begin{flushright} PURD-TH-95-03\\ hep-ph/9507380\\ July 1995\end{flushright}
\vspace{0.4cm}
\begin{center}
{\large\bf The Free Energy of Hot Gauge Theories with Fermions
Through\nolinebreak\ \boldmath $g^{\rm\bf 5}$\vspace{0.4cm}\vspace{0.8cm}\\}
Chengxing Zhai and Boris Kastening\vspace{0.8cm}\\
{\it Department of Physics\\ Purdue University\\ West Lafayette,
Indiana 47907-1396}\vspace{0.8cm}\vspace{0.4cm}\\
{\bf Abstract}\\
\end{center}

We compute the free energy density $F$ for gauge theories, with fermions,
at high temperature and zero chemical potential.
In the expansion $F=T^4 [c_0+c_2 g^2+c_3 g^3+(c'_4\ln g+c_4)g^4+
(c'_5\ln g+c_5)g^5+O(g^6)]$, we determine $c'_5$
and $c_5$ analytically by calculating two- and three-loop diagrams.
The $g^5$ term constitutes the first correction to the $g^3$ term
and is for the non-Abelian case the last power of $g$ that can be
computed within perturbation theory.
We find that the $g^5$ term receives no contributions from overlapping
double-frequency sums and that $c'_5$ vanishes.
\end{titlepage}

\section{Introduction}

The perturbative expansion of the free energy density of high-temperature
gauge theory in four dimensions can be written as
\begin{equation}
\label{fexp}
F=T^4 [c_0+c_2 g^2+c_3 g^3+(c'_4\ln g+c_4)g^4+(c'_5\ln g+c_5)g^5+O(g^6)],
\end{equation}
where the $c^{{\scriptscriptstyle (}_{'}{\scriptscriptstyle )}}_k$ are
numerical coefficients (that depend on the field content of the theory,
the renormalization scheme and the renormalization scale) and where we have
assumed the temperature high enough that fermion masses can be ignored.
Previously, $F$ has been computed to $O(g^3)$ for QED by Akhiezer
and Peletminskii \cite{AkPe} and for QCD by Kapusta \cite{Ka79},
while the $g^4\ln g$ term was obtained by Toimela \cite{To}.
More recently, $F$ has been computed to $O(g^4)$ by Arnold and one of the
present authors (C.Z.) \cite{ArZh}.
Corian\`o and Parwani \cite{PaCo} have recently studied
high-temperature QED up to $O(g^5)$.
The free energy density (or, equivalently, the pressure) is also known to
$O(g^5)$ in $\Phi^4$ theory (see Ref.\ \cite{PaSi} and references therein).
Here we determine the coefficients $c'_5$ and $c_5$ in expansion (\ref{fexp}).
For this purpose we need to take into account Debye screening at three
loops (for a review on Debye screening, see Refs.\ \cite{GrPiYa,Ka89}).

Note that, for the non-Abelian case, the $g^5$ term is believed to
be the last power in $g$ accessible within perturbation theory
\cite{Li} (for a review, see Refs.\ \cite{GrPiYa,Ka89}).
Starting at four loops, infrared problems that are believed to
be cured by non-perturbative magnetic screening lead to contributions
to the $g^6$ term from diagrams with arbitrarily high numbers of loops.

The $g^5$ term is also interesting because it constitutes the first
correction to the $g^3$ term, the lowest order at which Debye
screening plays a role.
The renormalization group invariance to this order can be tested.
The dependence of $F$ on the renormalization scale due to the $g^3$ term
should be diminished by including the $g^5$ term.
Checking this, we can gain some idea about the theoretical uncertainties
of the $g^3$ term as well as the behavior of the perturbative expansion.
Also, our result can be used for a test of an evaluation of $F$ on
the lattice.
Finally, our result is potentially interesting for the evolution of
the early Universe, where one might have to add scalars to the theory.

In section \ref{notation} notation and conventions are established.
In section \ref{procedure} we outline our general computational
procedure and emphasize what is new as compared to the $g^4$ calculation.
In section \ref{result} we conclude by presenting and analyzing
our result as well as comparing it to related results.

\section{Notation and Conventions}
\label{notation}
We use the same notation and conventions as in Ref.\ \cite{ArZh}.
We now present an almost verbatim review of these to keep this work as
self-contained as possible.

We consider gauge theories given in Euclidean spacetime by Lagrangians
of the form
\begin{equation}
{\cal L}_{\rm E} =
\bar{\psi}\gamma_\mu\left(\partial_\mu-igA_\mu^a T^a\right)\psi
+\frac{1}{4}\left(\partial_\mu A_\nu^a -\partial_\nu A_\mu^a
+gf^{abc}A_\mu^b A_\nu^c\right)^2+{\cal L}_{\rm gf,gh}\,,
\end{equation}
with gauge fixing and ghost term ${\cal L}_{\rm gf,gh}$, and where the
$T^a$ are the generators of a single, simple Lie group, such as U(1)
or SU(3).
To simplify our presentation, we will not derive results for an arbitrary
product of simple Lie groups such as SU(2)$\times$U(1), but such cases
could easily be handled by adjusting the overall group and coupling
factors on the results we give for individual diagrams.

$d_{\rm A}$ and $C_{\rm A}$ are the dimension and quadratic Casimir invariant
of the adjoint representation, with
\begin{equation}
\delta^{aa}=d_{\rm A}\,,\;\;\;\;\;\;\;\;\;\;\;\;\;\;\;
f^{abc}f^{dbc}=C_{\rm A}\delta^{ad}\,.
\end{equation}
$d_{\rm F}$ is the dimension of the total fermion representation
({\it e.g.}, 18 for six-flavor QCD), and $S_{\rm F}$ and $S_{2\rm F}$
are defined in terms of the generators $T^a$ for the total fermion
representation as
\begin{equation}
S_{\rm F} =\frac{1}{d_{\rm A}}{\rm tr}(T^2)\,,\;\;\;\;\;\;\;\;\;\;\;\;\;\;
S_{2\rm F} = \frac{1}{d_{\rm A}}{\rm tr} [(T^2)^2]\,,
\end{equation}
where $T^2 = T^a T^a$.
For SU($N$) with $n_{\rm f}$ fermions in the fundamental representation,
the standard normalization of the coupling gives
\begin{equation}
\label{SUNnotation}
d_{\rm A} = N^2-1 \,,
\;\;\;\;\;
C_{\rm A} = N \,,
\;\;\;\;\;
d_{\rm F} = N n_{\rm f} \,,
\;\;\;\;\;
S_{\rm F} = {1\over2} n_{\rm f} \,,
\;\;\;\;\;
S_{2\rm F} = {N^2{-}1 \over 4N} n_{\rm f} \,.
\end{equation}
For U(1) theory, relabel $g$ as $e$ and let the charges of the $n_{\rm f}$
fermions be $q_i e$.  Then
\begin{equation}
d_{\rm A} = 1 \,,\;\;\;\;\;
C_{\rm A} = 0 \,,\;\;\;\;\;
d_{\rm F} = n_{\rm f} \,,\;\;\;\;\;
S_{\rm F} = \sum_i q_i^2 \,,\;\;\;\;\;
S_{2\rm F} = \sum_i q_i^4 \,.
\label{QEDnotation}
\end{equation}
If the fermion representation is irreducible or consists of several
identical copies of an irreducible representation [as in
(\ref{SUNnotation}) above], we have
\begin{equation}
d_{\rm A}S_{\rm F}^2 = d_{\rm F}S_{2\rm F}\,.
\end{equation}

We work in Feynman gauge.
We also work exclusively in the Euclidean (imaginary time) formulation of
thermal field theory.
We conventionally refer to four-momenta with capital letters $K$ and
to their components with lower-case letters: $K=(k_0,\vec k)$.
All four-momenta are Euclidean with discrete frequencies $k_0 = 2\pi n T$
for bosons and ghosts and $k_0 = 2\pi \left(n{+}{1\over2}\right)T$ for
fermions.
We regularize the theory by working in $d=4{-}2\epsilon$ dimensions with the
modified minimal subtraction ($\overline{\rm MS}$) scheme, which
corresponds to doing minimal subtraction (MS) and then changing the MS
scale $\mu$ to the $\overline{\rm MS}$ scale $\bar\mu$ by the substitution
\begin{equation}
\mu^2 = \frac{e^{\gamma_{\rm\scriptscriptstyle E}}\bar{\mu}^2}{4\pi}\,.
\end{equation}
The trace over the identity in spinor space is by convention ${\rm tr} I=4$.

To denote summation over discrete loop frequencies and integration over
loop three-momenta, we use the shorthand notation
\begin{equation}
\hbox{$\sum$}\!\!\!\!\!\!\int_P \rightarrow
\mu^{2\epsilon}T\sum_{p_0}\int\frac{d^{3-2\epsilon}p}{(2\pi)^{3-2\epsilon}}
\end{equation}
for bosonic momenta and
\begin{equation}
\hbox{$\sum$}\!\!\!\!\!\!\int_{\{P\}} \rightarrow
\mu^{2\epsilon}T\sum_{\{p_0\}}\int\frac{d^{3-2\epsilon}p}{(2\pi)^{3-2\epsilon}}
\end{equation}
for fermionic momenta, where
\begin{equation}
\sum_{p_0}\rightarrow\sum_{p_0 = 2\pi n T}\,,\;\;\;\;\;\;\;\;\;\;
\sum_{\{p_0\}} \rightarrow \sum_{p_0 = 2\pi\left(n+\frac{1}{2}\right)T}\,.
\end{equation}

\begin{figure}
\epsfxsize=12.5cm
\centerline{\epsfbox{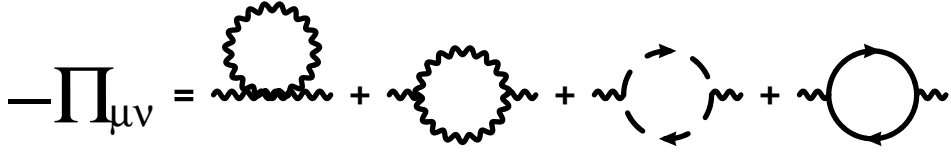}}
\caption{The one-loop gluon self-energy.}
\label {fig pi}
\end{figure}

We handle the resummation of hard thermal loops [which is required to
make perturbation theory well behaved beyond $O(g^2)$] as was done
in Ref.~\cite{ArZh}.
Specifically, we must improve our propagators by incorporating the
Debye screening mass $M$ for $A_0$, which is determined at leading order
by the self-energy diagrams of Fig.\ \ref{fig pi}:
\begin{equation}
\label{M def}
M^2\delta^{ab}=\Pi^{ab}_{00}(0)=\Pi^{ab}_{\mu\mu}(0)=g^2\delta^{ab}\left[
C_{\rm A} (d-2)^2 \hbox{$\sum$}\!\!\!\!\!\!\int_Q\frac{1}{Q^2}
-4S_{\rm F}(d-2)\hbox{$\sum$}\!\!\!\!\!\!\int_{\{Q\}}\frac{1}{Q^2}
\right]\,.
\end{equation}
This is accomplished by rewriting our Lagrangian density, in frequency
space, as
\begin{equation}
\label{reorganized L}
{\cal L}_{\rm E}=\left({\cal L}_{\rm E}+{\textstyle\frac{1}{2}}
M^2 A^a_0 A^a_0 \delta_{p_0^{}}\right)-{\textstyle\frac{1}{2}} M^2 A^a_0 A^a_0
\delta_{p_0^{}}\,,
\label {gauge resummation}
\end{equation}
where $\delta_{p_0^{}}$ is shorthand for the Kronecker delta symbol
$\delta_{p_0^{},0}$.
Then we absorb the first $A_0^2$ term into our unperturbed Lagrangian
${\cal L}_0$ and treat the second $A_0^2$ term as a perturbation.

Since the free energy density is computed by considering vacuum diagrams
(diagrams without external legs), there is no need to explicitly introduce
wave function renormalizations. We only need to renormalize the
coupling constant. We do this by expressing the bare coupling
constant $g_{\rm b}$ in terms of the renormalized coupling $g$,
\begin{equation}
\mu^{-2 \epsilon} g_{\rm b}^2 = Z_g^2 g^2
=\left(1-\frac{11C_{\rm A}-4S_{\rm F}}{3\epsilon}\frac{g^2}{(4\pi)^2}
+O(g^4)\right)g^2\,,
\end{equation}
and then using the bare coupling constant $g_{\rm b}$ to the required order
in $g$ at all vertices in the vacuum diagrams. Through $g^5$ it is
sufficient to know the one-loop renormalization given above.
However, for the computation of the Debye screening mass (\ref{M def})
we have used the renormalized coupling $g$. This is allowed because
for the cure of the infrared problems achieved by reorganizing the
perturbation expansion according to (\ref{reorganized L}), only the
leading contribution in $g$ to the Debye mass is crucial.

\section{Computational Procedure}
\label{procedure}

\subsection{\protect\boldmath $g^{\rm\bf 5}$ Order Contributions}

\begin{figure}
\epsfysize=16cm
\centerline{\epsfbox{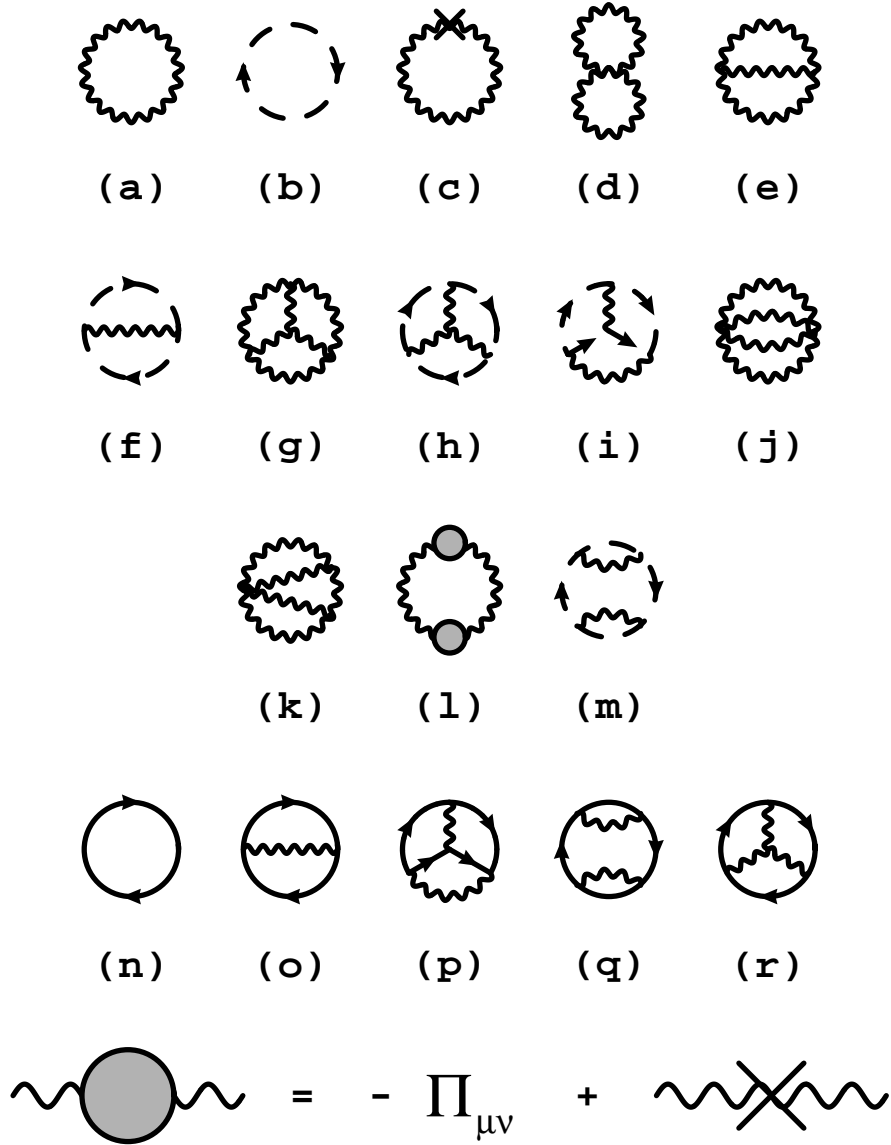}}
\caption{Diagrams contributing to the free energy of gauge theories with
fermions. The crosses are the ``thermal counterterms'' arising from the
last term of (\protect\ref{gauge resummation}).}
\label{diagrams}
\end{figure}

In the expansion of the free energy density, the zeroth-order term
represents the free energy density of an ideal gas containing
free gauge bosons and massless quarks.
The leading contribution due to the interaction is of order $g^2$
which is represented by two-loop diagrams.
For the calculation of higher-order terms the resummation
(\ref{gauge resummation}) for the static timelike gluon propagator
is required because of infrared divergences.
It is this resummation, which introduces the Debye screening mass into the
theory, that causes the expansion of the free energy density being
in powers of $g$ instead of $g^2$.
Consequently, one cannot determine the order of a diagram by naively
counting the number of interaction vertices.
The leading odd-power contribution is of order $g^3$ which comes from the
one-loop diagram with the resummed static gluon propagator.
The $g^4$ term receives contributions from the subleading pieces of
two-loop diagrams as well as the leading pieces of three-loop diagrams.
To get the $g^5$ term of the free energy density,
we need to compute the two-loop diagrams to higher order and
the subleading pieces of three-loop diagrams.
Fig.\ \ref{diagrams} contains all the diagrams contributing to the
free energy density up to $g^5$ order.

We mainly follow the formal manipulations in Refs.\
\cite{ArZh} to simplify the sum-integrals obtained by applying
the Feynman rules to the diagrams in Fig.\ \ref{diagrams}.
We shall focus on the contributions at order $g^5$.
Due to the resummation, there are two momentum scales,
$gT$ (the Debye mass) and $T$, appearing in the sum-integrals.
We will conveniently refer to momenta of order $gT$ as ``soft''
and momenta of order $T$ as ``hard.''
A sum-integral usually gains its dominant piece from a
momentum integral region where some momenta are hard and
others are soft.
Our strategy is to identify the soft and hard momenta
in the sum-integrals and carry out an expansion about the ratio
between the Debye mass or the soft momenta and the hard momenta
to extract the leading piece. We then construct a new sum-integral
from the original one by subtracting its leading piece out.
We then find the corresponding soft and hard momenta for
this new sum-integral to extract its leading piece which
is the next-to-leading piece for the original integral.
This procedure enables us to get a systematic expansion
in powers of $g$. Since when getting the leading piece of
a sum-integral, a soft momentum is always neglected compared
to a hard momentum, the integrals for two different scales $T$
and $gT$ are separated or decoupled.

In the following two subsections, we give more details about our
computational procedure by considering separately two-
and three-loop diagrams. We concentrate on explaining the way in
which we perform the calculation but spare the reader from
all the messy details for computing individual diagrams.
The expressions for all the diagrams in Fig.\ \ref{diagrams}
are provided in Appendix \ref{single diagrams}.

\subsection{Two-loop Diagrams}

To illustrate the general discussion in the previous subsection,
let us first consider a typical two-loop sum-integral
arising from the setting sun diagram (e) in Fig.\ \ref{diagrams},
\begin{equation}
    A_{\rm typ} \equiv \hbox{$\sum$}\!\!\!\!\!\!\int_{PQ}
	{\delta_{p_0} \over P^2 + M^2} {q_0^2 \over Q^2 (P+Q)^2} \,.
\label{A_resum_def}
\end{equation}
We shall use superscripts to denote the pieces in the expansion
of $A_{\rm typ}$ in $M$, {\it i.e.}, $A_{\rm typ}^{(0)}$ for the leading
piece, $A_{\rm typ}^{(1)}$ for the subleading piece and so on.
The leading piece of $A_{\rm typ}^{(0)}$ may be obtained by setting $M$ to
zero since $A_{\rm typ}$ is infrared safe as $M \to 0$ and since the typical
$P$ contributing to $A_{\rm typ}$ is of order $T$ which is much larger
than $M$.
We can then subtract this leading piece from $A_{\rm typ}$ which gives
\begin{equation}
    A_{\rm typ} = A_{\rm typ}^{(0)} - M^2 \hbox{$\sum$}\!\!\!\!\!\!\int_{PQ}
	{\delta_{p_0} \over P^2 (P^2 + M^2)} {q_0^2 \over Q^2 (P+Q)^2}
\label{leading}
\end{equation}
with
\begin{equation}
A_{\rm typ}^{(0)}=\hbox{$\sum$}\!\!\!\!\!\!\int_{PQ}{\delta_{p_0}\over P^2}
{q_0^2 \over Q^2 (P+Q)^2} \,.
\end{equation}
Now, we find that the $P$ integral in the second term on the right-hand
side of Eq.\ (\ref{leading}) is infrared sensitive to $M$ which means
this $P$ integral picks up contribution mainly at the region where $P$
is of order $M$.
Thus, $P$ is soft for the subleading piece of $A_{\rm typ}$.
Since $Q$ is always hard, the subleading piece $A_{\rm typ}^{(1)}$ is
\begin{equation}
    A_{\rm typ}^{(1)} = - M^2 \hbox{$\sum$}\!\!\!\!\!\!\int_{PQ}
        {\delta_{p_0} \over P^2 (P^2 + M^2)} {q_0^2 \over Q^4}
        = J_{\rm 1a}\;\hbox{$\sum$}\!\!\!\!\!\!\int_Q\frac{q_0^2}{Q^4}
\end{equation}
and then
\begin{equation}
\label{atyp01}
A_{\rm typ} = A_{\rm typ}^{(0)} + A_{\rm typ}^{(1)}
+ M^2 \hbox{$\sum$}\!\!\!\!\!\!\int_{PQ} {\delta_{p_0} \over P^2 (P^2 + M^2)}
	\left [{q_0^2 \over Q^4} - {q_0^2 \over Q^2 (P+Q)^2} \right ] \,.
\end{equation}
Here we have defined the integral
\begin{equation}
J_{\rm 1a}=\hbox{$\sum$}\!\!\!\!\!\!\int_P\delta_{p_0}\left(\frac{1}{P^2+M^2}
-\frac{1}{P^2}\right)
=-M^2\hbox{$\sum$}\!\!\!\!\!\!\int_P\frac{\delta_{p_0}}{P^2(P^2+M^2)}
\end{equation}
which is of order $g$ in four spacetime dimensions. Its value can be found
in Appendix \ref{basic integrals}.

Now since the $P$ integral in the last term of (\ref{atyp01}) behaves
as $1/(P^2+M^2)$ when $P$ is much less than $T$, we find that the $P$
integral receives its main contribution from the region where $P$ is
hard. Thus,
\begin{equation}
A_{\rm typ}^{(2)} = M^2 \hbox{$\sum$}\!\!\!\!\!\!\int_{PQ}
{\delta_{p_0} \over P^4}
\left [{q_0^2 \over Q^4} - {q_0^2 \over Q^2 (P+Q)^2} \right ]
\end{equation}
and
\begin{equation}
A_{\rm typ} = A_{\rm typ}^{(0)} + A_{\rm typ}^{(1)} + A_{\rm typ}^{(2)}
-M^4\hbox{$\sum$}\!\!\!\!\!\!\int_{PQ}
{\delta_{p_0} \over P^4 (P^2{+}M^2)}
\left [{q_0^2 \over Q^4}{-}{q_0^2 \over Q^2 (P{+}Q)^2} \right ] \,.
\end{equation}
We then identify $P$ as soft to get $A_{\rm typ}^{(3)}$ as
\begin{equation}
    A_{\rm typ}^{(3)} = - M^4 \hbox{$\sum$}\!\!\!\!\!\!\int_{PQ}
{\delta_{p_0} \over P^4 (P^2 + M^2)}
q_0^2 \left [{P^2 \over Q^6} - {4 (P \cdot Q)^2 \over Q^8}
\right ]
=\frac{1}{d{-}1}M^2 J_{\rm 1a}\;\hbox{$\sum$}\!\!\!\!\!\!\int_Q
\frac{q_0^2}{Q^6}
 \,,
\end{equation}
where we have expanded the denominator $(P+Q)^2$ in powers of
the ratio of $P$ and $Q$, replaced $\delta_{p_0}(P\cdot Q)^2$ by
$\delta_{p_0} p^2 q^2/(d-1)$ and integrated by parts in $q$.
It is not hard to see that $A_{\rm typ}^{(0)}$, $A_{\rm typ}^{(1)}$,
$A_{\rm typ}^{(2)}$, and $A_{\rm typ}^{(3)}$ are of orders $1$, $g$,
$g^2$, and $g^3$, respectively, in four spacetime dimensions.

For a two-loop diagram, there is an extra factor
$g_{\rm b}^2= \mu^{-2\epsilon} Z_g^2g^2$
multiplying $A_{\rm typ}$.
Thus, $A_{\rm typ}^{(1)}$ and $A_{\rm typ}^{(3)}$ contribute to the $g^5$
part of $F$. Therefore, the above asymptotic expansion of $A_{\rm typ}$
demonstrates how we extract the $g^5$ order contributions to the free energy
density from two-loop diagrams.

\subsection{Three-loop Diagrams}

We now turn to the three-loop diagrams.
Previously \cite{ArZh}, the Debye mass in the resummed propagator
(\ref{gauge resummation}) has been ignored for the three-loop
diagrams since after reorganizing perturbation theory the
three-loop diagrams are infrared finite if we set the Debye mass
to zero. Now, since we need to explore one more order, the
subleading terms need to be extracted. Since the Debye mass
only appears in the static gluon propagator and is only probed by
soft momenta of order $gT$, the sum-integrals that we need to deal
with contain at most two frequency sums and at least one soft-momentum
integral.

Three different cases appear:

(1) Case one is where we have pure three-dimensional triple-momentum
integrals with the Debye mass being the only mass scale.
Since there are three loops, there is a prefactor $g^4 T^3$ for these
triple-momentum integrals.
Thus, these three-dimensional triple-momentum integrals will give a result
proportional to the Debye mass to make up for the missing mass dimension,
{\it i.e.}, they contribute to the free energy density at order
$g^4 T^3 M \propto g^5 T^4$.
These pure three-dimensional three-loop integrals are one of the new
features which we encounter in the $g^5$ order calculation.
Their evaluations are provided in Appendix \ref{3-dim evaluations}.

(2) The second case we consider is where the sum-integral contains only
one sum and two three-dimensional soft-momentum integrals.
Thus, in this case there is only one hard loop momentum integral.
Neglecting the soft momentum relative to the hard momentum enables us
to decouple the three-loop integral into a product of a one-loop
sum-integral and a two-loop pure three-dimensional momentum integral,
which can be evaluated by standard methods.
In fact, we can show that these do not contribute to the free energy
density at order $g^5$.
The point is that a two-loop three-dimensional momentum integral
produces a result proportional to the Debye mass to an even power.
Since only soft-momentum integrals generate odd powers in $g$, this
case is not relevant to the $g^5$ order evaluation.

(3) Finally, we need to consider the case where two sum-integrals and
one soft-momentum integral are involved.
Again, for the hard-momentum sum-integral, it is valid to neglect the
soft momentum compared to the hard one.
This leads to a product of a two-loop sum-integral and a one-loop
three-dimensional momentum integral.
With the methods developed in Ref.\ \cite{ArZh}, we can evaluate
the two-loop sum-integrals.
However, it turns out that after we sum up all the pieces contributing
to free energy density at $g^5$ order, all the overlapping double
sum-integrals cancel and only the non-overlapping double
sum-integrals, which can be written as a product of two one-loop
sum-integrals, survive. This observation was already made earlier
for the case of QED \cite{PaCo}.

As a concrete example, let us consider the simplest
three-loop diagram, the basketball diagram (j) in Fig.\ \ref{diagrams}.
Applying the Feynman rules gives
\begin{eqnarray}
    - \mu^{2 \epsilon} F_j &=& {3 \over 16} d(d{-}1)
	g^4 d_{\rm A} C_{\rm A}^2 I_{\rm ball}^{\rm bb}
	+ {3(d{-}1) \over 8} g^4 d_{\rm A} C_{\rm A}^2
	\hbox{$\sum$}\!\!\!\!\!\!\int_{PQK}
\Biggr [\!\left({1{-}\delta_{p_0} \over P^2} +
	{\delta_{p_0} \over P^2{+}M^2} \right )
\nonumber\\
	&& \quad \times \left ({1{-}\delta_{q_0} \over Q^2} +
	{\delta_{q_0} \over Q^2{+}M^2} \right )
	- {1 \over P^2 Q^2} \Biggr ] {1 \over K^2 (P{+}Q{+}K)^2} \,.
\end{eqnarray}
The definition of $I_{\rm ball}^{\rm bb}$ may be found in
Appendix \ref{basic integrals}.
It is convenient to rewrite the expression above as
\begin{eqnarray}
\label{Fj}
-\mu^{2 \epsilon} F_j = {3 \over 8} (d-1) g^4 d_{\rm A} C_{\rm A}^2
&& \Biggr \{ {1 \over 2} \, d \, I_{\rm ball}^{\rm bb}
\nonumber\\
&&
+ \hbox{$\sum$}\!\!\!\!\!\!\int_{PQK} \!
\delta_{p_0}\delta_{q_0}\delta_{k_0}
\left [{1 \over (P^2{+}M^2)(Q^2{+}M^2)}{-}{1 \over P^2 Q^2}
\right ]{1 \over K^2 (P{+}Q{+}K)^2}
\nonumber\\
&& + \hbox{$\sum$}\!\!\!\!\!\!\int_{PQK} \left ({\delta_{p_0}
\over P^2{+}M^2}{-}{\delta_{p_0} \over P^2} \right )
\left ({\delta_{q_0}
\over Q^2{+}M^2}{-}{\delta_{q_0} \over Q^2} \right )
{1{-}\delta_{k_0} \over K^2 (P{+}Q{+}K)^2}
\nonumber\\
&& + 2 \hbox{$\sum$}\!\!\!\!\!\!\int_{PQK} \left ({\delta_{p_0}
\over P^2{+}M^2}{-}{\delta_{p_0} \over P^2} \right )
{1-\delta_{q_0}\delta_{k_0} \over Q^2 K^2 (P{+}Q{+}K)^2}\Biggr \} \,.
\label{basketball}
\end{eqnarray}
Let us consider each term at the right-hand side of
Eq.\ (\ref{basketball}) above.
The first term, involving $I_{\rm ball}^{\rm bb}$, is of order $g^4$ and
represents the leading contribution of many three-loop diagrams and has
been evaluated in Ref.\ \cite{ArZh}.

Recall that we listed three cases in the general treatment of the
three-loop diagrams. The second, third, and fourth terms correspond
to these three cases, respectively:

The second term is a three-loop pure three-dimensional momentum integral.
This is the first case discussed above.
We encounter a class of these three-dimensional integrals which are
defined in Appendix \ref{basic integrals} and computed in
Appendix \ref{3-dim evaluations}.

The third term involves two soft-momentum integrals corresponding
to the second case. Since $K$ is hard,
it is valid to neglect $P,Q$ compared to $K$ to write this third
term as
\begin{displaymath}
\frac{3(d-1)}{8}g^4d_{\rm A}C_{\rm A}^2\hbox{$\sum$}
\!\!\!\!\!\!\int_{PQ}\delta_{p_0}\delta_{q_0}
	\left (\frac{1}{P^2{+}M^2}-\frac{1}{P^2}\right)
	\left (\frac{1}{Q^2{+}M^2}-\frac{1}{Q^2}\right)
	\hbox{$\sum$}\!\!\!\!\!\!\int_K \frac{1}{K^4}
\end{displaymath}
\begin{equation}
=\frac{3(d-1)}{8}g^4d_{\rm A} C_{\rm A}^2 J_{\rm 1a}^2\;\hbox{$\sum$}
\!\!\!\!\!\!\int_K\frac{1}{K^4}
\end{equation}
which is of order $g^6$ as what we have expected (even power in $g$).

The fourth term corresponds to the third case where $P$ is
the soft momentum since $P$ needs to resolve the Debye mass $M$.
Therefore, as described above, we can approximate $(P+Q+K)^2$
as $(Q+K)^2$ (the case where $q_0+k_0=0$ and $|\vec{q}+\vec{k}|$ is soft
contributes only at order $g^6$ because of phase space suppression).
Now, the $P$ integral decouples from the $Q,K$ sum-integral.
Therefore the fourth term in (\ref{Fj}) can be expressed as a product
of a single three-dimensional momentum integral and a double sum-integral:
\begin{equation}
    {3(d-1) \over 4} g^4 d_{\rm A} C_{\rm A}^2 J_{\rm 1a}
	I_{\rm sun}^b \,,
\end{equation}
where we have introduced $I_{\rm sun}^b$ as
\begin{equation}
\label{b_sun_def}
    I_{\rm sun}^b \equiv \hbox{$\sum$}\!\!\!\!\!\!\int_{QK}
{1 \over Q^2 K^2 (Q+K)^2}
\end{equation}
and omitted the term vanishing in dimensional regularization.

Therefore, we have explicitly shown how to extract the order $g^5$
contribution to the free energy density from a simple three-loop diagram.
These order $g^5$ contributions are expressed as either
three-loop pure three-dimensional momentum integrals or as products
of a single three-dimensional soft-momentum integral and a double
sum-integral.
For other three-loop diagrams, parallel steps can be followed except
for possibly more elaborate expressions; of course, we need to introduce
more double sum-integrals and pure three-dimensional momentum integrals.
However, as mentioned before, when we add up all the pieces contributing
to the free energy at order $g^5$ from the three-loop diagrams,
these double sum-integrals cancel except for the ``non-overlapping''
sum-integrals that can be expressed as a product of two single
sum-integrals. We do not know a fundamental reason which leads to this
cancellation.

\section{Result and Analysis}
\label{result}
Combining the results for all the diagrams listed in Appendix
\ref{single diagrams} produces the final result for the free energy
density through order $g^5$ in four spacetime dimensions as
\begin{eqnarray}
F &=& d_{\rm A} T^4\, {\pi^2\over9} \Biggr \{
- {1 \over 5} \left(1 + {7d_{\rm F}\over4d_{\rm A}} \right)
+ \left ({g \over 4 \pi} \right )^2
\left(C_{\rm A}+\textstyle{5\over2}S_{\rm F}\right)
\nonumber\\
&& \qquad
- 48 \left ({g \over 4 \pi} \right )^3
\left(\frac{C_{\rm A}+S_{\rm F}}{3}\right)^{3/2}
- 48 \left ({g \over 4 \pi} \right )^4 C_{\rm A} (C_{\rm A}+S_{\rm F})
\ln\left({g\over 2\pi}\sqrt{C_{\rm A}+S_{\rm F} \over 3}\right)
\nonumber\\
&& \qquad
+ \left(g \over 4 \pi\right )^4 \Biggr [
C_{\rm A}^2
\left(
{22 \over 3} \ln\frac{\bar\mu}{4\pi T}
{+}{38 \over 3} \frac{\zeta'(-3)}{\zeta(-3)}
{-}{148 \over 3} \frac{\zeta'(-1)}{\zeta(-1)}
{-}4 \gamma_{\rm\scriptscriptstyle E}
{+}{64 \over 5}
\right)
\nonumber\\
&& \qquad\qquad\qquad\;
+C_{\rm A}S_{\rm F}
\left(
{47 \over 3} \ln\frac{\bar\mu}{4\pi T}
{+}{1 \over 3} \frac{\zeta'(-3)}{\zeta(-3)}
{-}{74 \over 3} \frac{\zeta'(-1)}{\zeta(-1)}
{-}8 \gamma_{\rm\scriptscriptstyle E}
{+}{1759 \over 60}
{+}{37\over5}\ln 2
\right)
\nonumber\\
&& \qquad\qquad\qquad\;
+S_{\rm F}^2
\left(
{-}{20 \over 3} \ln\frac{\bar\mu}{4\pi T}
{+}{8 \over 3} \frac{\zeta'(-3)}{\zeta(-3)}
{-}{16 \over 3} \frac{\zeta'(-1)}{\zeta(-1)}
{-}4 \gamma_{\rm\scriptscriptstyle E}
{-}{1 \over 3}
{+}{88\over5}\ln 2
\right)
\nonumber\\
&& \qquad\qquad\qquad\;
+S_{2\rm F}
\left(
{-}{105 \over 4}
{+}24\ln 2
 \right)
\Biggr ]
\nonumber\\
&& \qquad
- \left (g \over 4 \pi \right )^5
\left (C_{\rm A}{+}S_{\rm F} \over 3 \right)^{1/2} \Biggr [C_{\rm A}^2
\left ( 176 \ln\frac{\bar\mu}{4\pi T}
{+}176 \gamma_{\rm\scriptscriptstyle E}
{-}24 \pi^2
{-}494
{+}264 \ln 2
\right )
\nonumber\\
&& \qquad \qquad \qquad \qquad \qquad \qquad\;
+ C_{\rm A} S_{\rm F} \left (112 \ln\frac{\bar\mu}{4\pi T}
{+}112 \gamma_{\rm\scriptscriptstyle E}
{+}72
{-}128 \ln 2
\right )
\nonumber\\
&& \qquad \qquad \qquad \qquad \qquad \qquad\;
+ S_{\rm F}^2 \left (-64 \ln\frac{\bar\mu}{4\pi T}
{-}64 \gamma_{\rm\scriptscriptstyle E}
{+}32
{-}128 \ln 2
\right )
\nonumber\\
&& \qquad \qquad \qquad \qquad \qquad \qquad\;
- 144 S_{2\rm F}
\Biggr ]
+ O(g^6) \Biggr \} \,,
\end{eqnarray}
where $\zeta$ is Riemann's zeta function and
$\gamma_{\rm\scriptscriptstyle E}$ is the Euler-Mascheroni constant.

For QCD with $n_f$ quark flavors, to $g^5$ order, the free
energy density is
\begin{eqnarray}
    F = - && {8 \pi^2 T^4\over 45} \biggl\{
     1 + \textstyle{21\over32}n_{\rm f}
     - 0.09499\, g^2 \left(1 + \textstyle{5\over12}n_{\rm f}\right)
     + 0.12094\, g^3 \left(1+\textstyle{1\over6}n_{\rm f}\right)^{3/2}
\nonumber\\
&&
     + g^4 \biggl[
         0.08662 \left(1+\textstyle{1\over6}n_{\rm f}\right)
                     \ln\left(g\sqrt{1+\textstyle{1\over6}n_{\rm f}}\right)
         - 0.01323 \left(1+\textstyle{5\over12}n_{\rm f}\right)
                   \left(1-\textstyle{2\over33}n_{\rm f}\right)
                   \ln{\bar\mu\over T}
\nonumber\\
&& \qquad
         + 0.01733 - 0.00763\,n_{\rm f} - 0.00088\,n_{\rm f}^2
     \biggr]
\nonumber\\
&&
    + g^5 \sqrt{1+\textstyle{1\over6}n_{\rm f}} \biggl[
	0.02527 \left(1+\textstyle{1\over6}n_{\rm f}\right)
                   \left(1-\textstyle{2\over33}n_{\rm f}\right)
                   \ln{\bar\mu\over T}
\nonumber\\
	&& \qquad \qquad \qquad
	-0.12806 - 0.00717 \, n_f + 0.00027 \, n_f^2 \biggl ]
	+ O(g^6) \biggl \} \,,
\end{eqnarray}
where we have evaluated the coefficients numerically.

For QED with $n_f$ charged fermions with charges $q_i e$, the
fifth-order free energy density may be read from the expression
above on using~(\ref{QEDnotation})
\begin{equation}
    F_{\rm QED}^{(5)} = {\pi^2 T^4 \over 9 \sqrt{3}}
	\left (\sum q_i^2 \right )^{1/2} \left (e \over 4 \pi \right )^5
	\left [ \left (\sum q_i^2 \right )^2
	\left (64 \ln\frac{\bar\mu}{4\pi T}
        + 64 \gamma_{\rm\scriptscriptstyle E} - 32 + 128 \ln 2 \right )
	+ 144 \sum q_i^4 \right ] \,.
\end{equation}
Taking $\mu=T$, it is not hard to check that this result agrees with
Ref.\ \cite{PaCo}.

As in Ref.\ \cite{ArZh}, we now check whether the perturbative
expansion of the QCD free energy density behaves well for
physically realized values of couplings to $g^5$ order.
Although the free energy does not have any renormalization scale
$\mu$ dependence, the partial sum does so.
If the perturbative expansion is well behaved, including
higher-order corrections into the partial sum reduces the $\mu$
dependence.
The inclusion of the $g^5$ order term in the partial sum compensates the
$\mu$ dependence of the $g^3$
term\footnote{In Ref.\ \cite{ArZh}, with the help of the renormalization
group a $g^5\ln\bar\mu/T$ was introduced to compensate for the $\mu$
dependence due to the $g^3$ term. We note that our present $g^5$ order
result correctly produces this desired $g^5\ln\bar\mu/T$ term.}.
Besides looking at the $\mu$ dependence of the partial sums, we also
compare the size of the contributions from each order.

\begin{figure}[t]
\epsfxsize=12.5cm
\centerline{\epsfbox{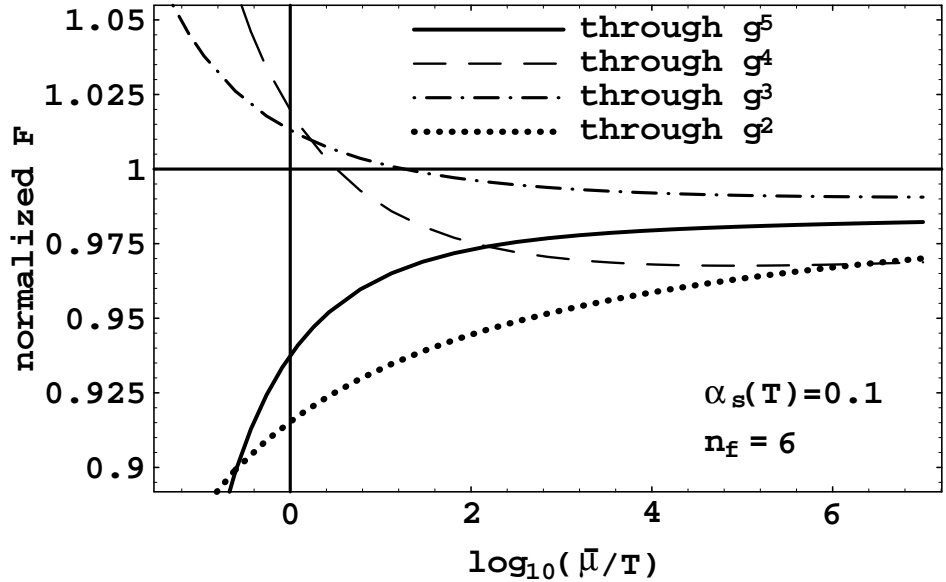}}
\caption{The dependence of the free energy density $F$ on the choice of
	renormalization scale $\bar\mu$ for six-flavor QCD with
        $\alpha_s(T) = 0.1$.
        The free energy density is normalized in units of the ideal gas
        result $-({1\over45}d_{\rm A} + {7\over180}d_{\rm F}) \pi^2 T^4$.
        The dotted, dot-dashed, dashed, and solid lines are the
	results for $F$ including terms through orders $g^2$, $g^3$,
	$g^4$, and $g^5$, respectively.}
\label {fig53a}
\end{figure}

Define $\alpha_s(T)\equiv g^2(T)/(4\pi)$.
Fig.\ \ref{fig53a} shows the result for six-flavor QCD when $\alpha_s(T){=}0.1$
(which corresponds to scales of order a few 100 GeV). The free energy density
is plotted vs the choice of renormalization scale $\bar\mu$.
We have taken
\begin{equation}
   {1\over g^2(\bar\mu)} \approx
      {1\over g^2(T)} - \beta_0 \ln{\bar\mu\over T}
      + {\beta_1\over\beta_0}
         \ln\left(1 - \beta_0 g^2(T) \ln{\bar\mu\over T}\right)
      \,,
\end{equation}
where
\begin{equation}
   \beta_0 = {1\over(4 \pi)^2}
\left(-\textstyle{22\over3}C_{\rm A}+\textstyle{8\over3}S_{\rm F}\right)\,,
   \qquad
   \beta_1 = {1\over(4 \pi)^4}
\left(-\textstyle{68\over3}C_{\rm A}^2+\textstyle{40\over3}C_{\rm A}S_{\rm F}
           + 8S_{2\rm F} \right) \,.
\end{equation}
In Ref.~\cite{ArZh}, it was found that including the $g^4$ term does not
make the partial sum for the free energy density less dependent on the
renormalization scale. There, one of the main sources of the $\mu$
dependence is the $g^3$ term which requires the order $g^5$ term to balance
its renormalization scale dependence. According to Fig.~\ref{fig53a},
inclusion of the $g^5$ term in the partial sum does
not generally make this sum less dependent on $\mu$
and the perturbative expansion does not behave well in this respect.
For $\bar\mu=T$, the terms at each order are
\begin{equation}
    F =  - {79 \pi^2 T^4 \over 90} \left [
	1 - 0.0846 + 0.0976 + (0.0255 + 0 - 0.0192) +
	(0 - 0.0818) + O (g^6) \right ] \,.
\label{qcdT}
\end{equation}
For this value of $\alpha_s$, the $g^2$ and $g^3$ terms have about the same
size. This does not necessarily mean that perturbation theory does not
work well since the $g^3$ term is the leading term of new physics at the
scale of $gT$ instead of being a correction to the $g^2$ term. If the
corrections at $g^4$ and $g^5$ are smaller than the $g^2$ and $g^3$ terms,
perturbation theory may still work well. However, the numerical values
above show that the $g^5$ term appears not to be generally smaller than
the $g^3$ term. Therefore, perturbation theory seems not to work well
for this value of $\alpha_s$ which corresponds to QCD at the electroweak
scale.

\begin{figure}
\epsfxsize=12.5cm
\centerline{\epsfbox{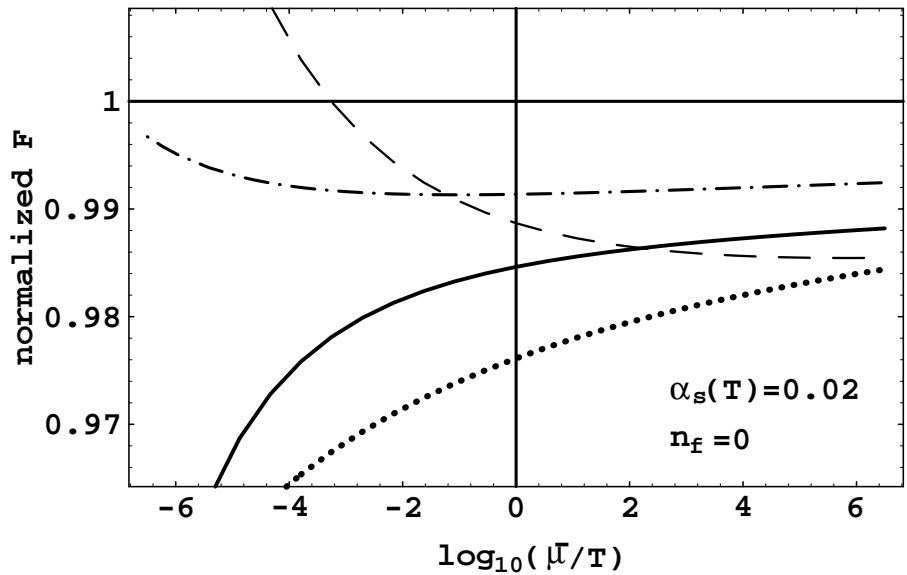}}
\caption{The same as Fig.\ \protect\ref{fig53a} but for
	$\alpha_s(T) = 0.02$ without fermions.}
\label {fig53b}
\end{figure}

Fig.~\ref{fig53b} shows the $\mu$ dependence for $\alpha_s(T)=0.02$ without
any fermions, {\it i.e.}, with $n_f = 0$.
This is interesting since it is [see Eq.\ (\ref{SUNnotation})]
equivalent to pure SU(2), {\it i.e.}, electroweak gauge theory at the
electroweak scale, with $\alpha_{\rm w} \approx 1/33$.
It is not hard to see that the $g^5$ order free energy density is less
sensitive to the renormalization scale than the $g^4$ order free energy
density.
However, the free energy density through $g^5$ order is not more stable
than the result through $g^3$ order.
This is due to a large cancellation between the $g^3$ term and the $g^2$
term.
Here are the values for the contributions at each order to the free
energy density for the choice $\bar\mu=T$:
\begin{equation}
    F = - { d_{\rm A} \pi^2 T^4 \over 45} \left [
	1 - 0.0239 + 0.0152 + (-0.00378 + 0 + 0.00109)
	+ (0 - 0.00406) + O (g^6) \right ] \,.
\label{ewT}
\end{equation}
Obviously, the corrections at orders $g^4$ and $g^5$ are smaller
than the $g^2$ and $g^3$ terms. This suggests that perturbation theory
works.

\begin{figure}
\epsfxsize=12.5cm
\centerline{\epsfbox{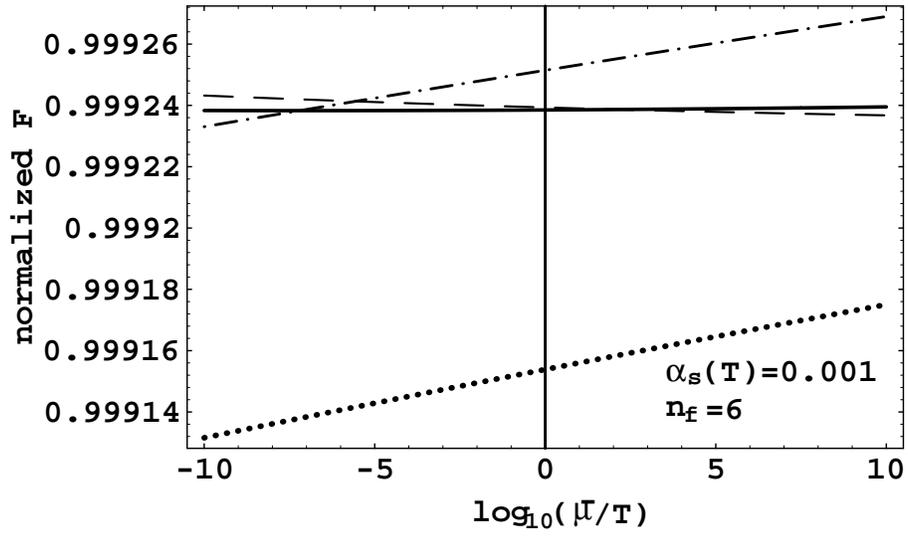}}
\caption{The same as Fig.\ \protect\ref{fig53a} but for
	$\alpha_s(T) = 0.001$.}
\label {fig53c}
\end{figure}

\begin{figure}
\epsfxsize=12.5cm
\centerline{\epsfbox{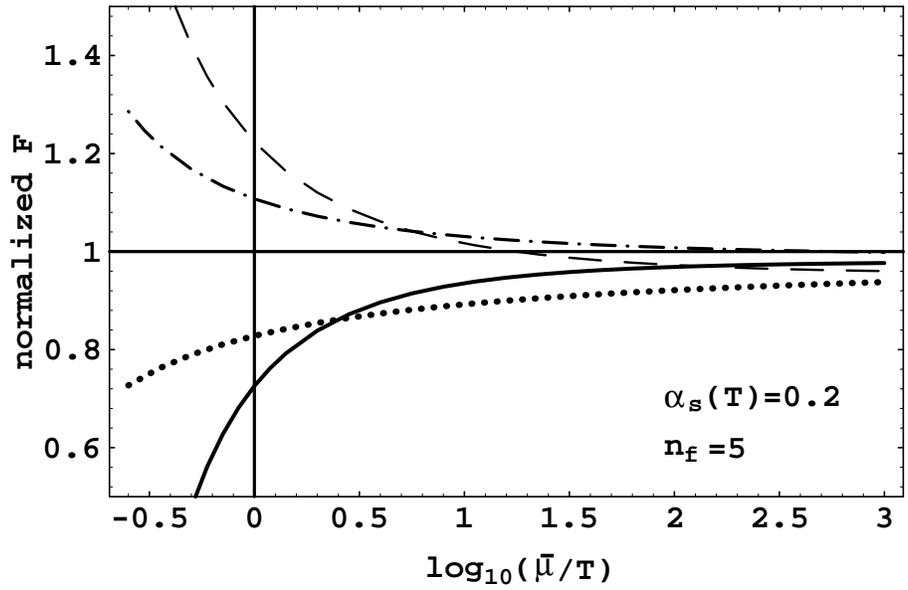}}
\caption{The same as Fig.\ \protect\ref{fig53a} but for
	$\alpha_s(T) = 0.2$ and $n_f = 5$.}
\label {fig53d}
\end{figure}

In Fig.~\ref{fig53c} we show a similar plot for $n_f =6$ and
$\alpha_s(T)=0.001$ where the behavior of the perturbative expansion is good.
In Fig.\ \ref{fig53d}, we provide the corresponding plot for
$\alpha_s (T) = 0.2$ but $n_f=5$ (for $T$ being several GeV).

We like to comment on the absence of the $g^5 \ln g$ term in
the expansion of the free energy density.
It is convenient to view the contributions to the free energy at each
order with an effective field theory technique as was done in
Refs.\ \cite{BrNi1}. In hot gauge theories, there are three relevant scales
in the imaginary time formalism: $T$, $gT$, and $g^2 T$. The scale $T$ is
related to the nonstatic fields while $gT$ is the scale for the Debye
screening effect. The scale $g^2 T$ is believed to be the inverse of the
magnetic screening length which cures the remaining infrared problem of
hot non-Abelian gauge theories. Since the $g^2T$ scale contributes to the
free energy starting only at order $g^6$, we can ignore it.
Imagine first integrating out the nonstatic fields (scale $T$
physics) to arrive at an effective field theory which correctly
describes physics in the low energy region (of order $gT$).
Let $\Lambda$ be the cutoff separating the scales $T$ and $gT$.
Integrating out the nonstatic fields gives a contribution to the free
energy density $f (T,\Lambda, g^2)$ which has an even power expansion
in $g$ since no resummation is required. Integrating out these nonstatic
fields also generates effective interaction terms for the static fields.
This introduces a dependence on the cutoff $\Lambda$ into the bare parameters
of the effective field theory for the static fields at scale $gT$ with
cutoff $\Lambda$.
An $\ln g$ term arises only through the logarithm of the ratio of
the scales $gT$ and $T$, {\it i.e.}, through cancellation between
$\ln\Lambda / T$ and $\ln \Lambda/(gT)$.
$\ln \Lambda/T$ terms enter the free energy density through
$f(T,\Lambda,g^2)$ and the bare parameters of the effective theory
for the static field at scale $gT$. It can be shown that there
are only two parameters relevant to the free energy density through
$g^5$, the effective mass and the effective coupling constant \cite{BrNi1}.
Since the coupling constant in the (superrenormalizable) effective theory
requires no renormalization, there will be no $\ln \Lambda/T$ appearing
inside the effective coupling.
Since $f(T, \Lambda, g^2)$ has an even power expansion
in $g$, a $g^5 \ln g$ term comes only from the cancellation
between an $\ln \Lambda/T$ term in the effective mass term and
another $\ln \Lambda /(gT)$ coming from the evaluations of the
effective theory.
Therefore, the absence of the $g^5 \ln g$ term means that there is no
$\ln\Lambda /T$ at order $g^4$
in the effective mass term which has been examined
explicitly in Ref.\ \cite{Na}.
In other words, this effective mass has vanishing anomalous dimension
and does not ``run'' at the leading order as we vary the cutoff $\Lambda$
\cite{Brpc}.
In fact, at the next-to-leading order, the effective mass does ``run''
\cite{FaKaRuSh}.

As an outlook, it would be interesting to investigate the reasons
for the cancellation of overlapping double-frequency sums as well
as for the absence of a $g^5\ln g$ term in $F$.
It would further be worthwhile to include scalar fields and to
consider the case of non-vanishing chemical potential.

Note added: Recently, using a different method, Braaten and Nieto
have confirmed our result \cite{BrNi2}.

\bigskip

We thank T.~Clark and S.~Love for useful discussions and advice.
We are grateful to P.~Arnold for suggestions on analyzing the behavior
of perturbation theory. We also thank E.~Braaten and A.~Nieto for
beneficial discussions and communications.
This work was supported by the U.S. Department of Energy,
contract No.\ DE-FG02-91ER40681 (Task B).


\newpage
\appendix

\section{Results for individual diagrams}
\label{single diagrams}
Here are the contributions to the free energy density through
$g^5$ order from individual diagrams in Fig.\ \ref{diagrams}.
The newly appearing symbols are defined in Appendix
\ref{basic integrals}.
\begin{eqnarray}
-\mu^{2\epsilon}F_a
&=&
-\frac{1}{2}d_{\rm A} d\,\hbox{$\sum$}\!\!\!\!\!\!
\int_P\ln P^2-d_{\rm A}\frac{1}{d-1}M^2J_{\rm 1a}\\
-\mu^{2\epsilon}F_b
&=&
d_{\rm A}\hbox{$\sum$}\!\!\!\!\!\!\int_P\ln P^2\\
-\mu^{2\epsilon}F_c
&=&
\frac{1}{2}d_{\rm A} M^2 J_{\rm 1a}\\
-\mu^{2\epsilon}F_d
&=&
-\frac{Z_g^2g^2}{4}d_{\rm A}C_{\rm A}(d-1)(d b_1^2+2J_{\rm 1a} b_1)\\
-\mu^{2\epsilon}F_e
&=&
-\frac{Z_g^2g^2}{24}d_{\rm A}C_{\rm A}\{
-18(d-1)b_1^2
+6(d-1)(2d-7)J_{\rm 1a} b_1\nonumber\\
&&
\;\;\;\;+6[J_{\rm 1a}^2+4M^2 J_{\rm 2a}
+4M^2 A_1^{\rm b}-2(2d-3)M^2 A_2^{\rm b}]\nonumber\\
&&
\;\;\;\;+(2d^2-13d+39)M^2J_{\rm 1a} b_2\}+O(g^6)
\\
-\mu^{2\epsilon}F_f
&=&
-\frac{Z_g^2g^2}{24}d_{\rm A}C_{\rm A}[6b_1^2
-6(d-3)J_{\rm 1a} b_1+12M^2 A_2^{\rm b}
-(d-5)M^2J_{\rm 1a} b_2]+O(g^6)\\
-\mu^{2\epsilon}F_g
&=&
\frac{g^4}{32}d_{\rm A}C_{\rm A}^2\{(20d-23)I_{\rm ball}^{\rm bb}\nonumber\\
&&
\;\;\;\;+4J_{\rm 1a}[4I_{\rm sun}^{\rm b}+(16d-23)I_{\rm 2a}
-(16d-28)I_{\rm 2b}
-2(4d-7)(d-3)b_1b_2]\nonumber\\
&&
\;\;\;\;+2(3J_{\rm 3a}+4J_{\rm 3b}+16J_{\rm 3c}-36J_{\rm 3e}+16J_{\rm 3i}
+12J_{\rm 3j}-12J_{\rm 1a}J_{\rm 2a})\}
+O(g^6)\\
-\mu^{2\epsilon}F_h
&=&
\frac{g^4}{16}d_{\rm A}C_{\rm A}^2\{-I_{\rm ball}^{\rm bb}
+4J_{\rm 1a}[-I_{\rm 2a}+2I_{\rm 2b}+(d-3)b_1b_2]\}+O(g^6)\\
-\mu^{2\epsilon}F_i
&=&
-\frac{g^4}{32}d_{\rm A}C_{\rm A}^2(I_{\rm ball}^{\rm bb}
+4J_{\rm 1a} I_{\rm 2a})+O(g^6)\\
-\mu^{2\epsilon}F_j
&=&
\frac{3g^4}{16}d_{\rm A}C_{\rm A}^2(d-1)(dI_{\rm ball}^{\rm bb}
+4J_{\rm 1a} I_{\rm sun}^{\rm b}+2J_{\rm 3b})+O(g^6)\\
-\mu^{2\epsilon}F_k
&=&
\frac{3g^4}{16}d_{\rm A}C_{\rm A}^2\{-9(d-1)I_{\rm ball}^{\rm bb}
-12J_{\rm 1a}[I_{\rm sun}^{\rm b}+(2d-3)I_{\rm 2a}]\nonumber\\
&&
\;\;\;\;+4(-2J_{\rm 3b}+3J_{\rm 3e}-J_{\rm 3j}+J_{\rm 1a}J_{\rm 2a})\}
+O(g^6)\\
-\mu^{2\epsilon}F_l
&=&
\frac{1}{4}I_{qcd}
-\frac{g^4}{4}d_{\rm A}C_{\rm A}J_{\rm 1a}\nonumber\\
&&
\;\;\;\times\{C_{\rm A}[(3d-10)I_{\rm sun}^{\rm b}-(10d-12)I_{\rm 2a}
-8(d-2)I_{\rm 2b}-2(d-2)(d^2-6d+6)b_1b_2]\nonumber\\
&&
\;\;\;\;\;\;\;\;+S_{\rm F}[-4(d-4)I_{\rm sun}^{\rm f}+16(d-3)I_{\rm 2c}
+32I_{\rm 2d}+8(d^2-6d+6)f_1b_2]\}\nonumber\\
&&
+\frac{g^4}{16}d_{\rm A}C_{\rm A}^2\left[
-2(d-3)\frac{J_{\rm 1a}^3}{M^2}+4J_{\rm 1a}\left(-\frac{3d-5}{d-2}J_{\rm 2a}
+4\frac{d-3}{d-2}J_{\rm 2c}+8J_{\rm 2d}\right)\right.\nonumber\\
&&
\;\;\;\;\;\;\;\;\;\;\;\;\;\;\;\;\;\;
+\left(\frac{1}{d-2}J_{\rm 3a}+2\frac{5d-11}{d-2}J_{\rm 3b}
+\frac{8}{d-2}J_{\rm 3c}-64J_{\rm 3e}+64J_{\rm 3f}\right.\nonumber\\
&&
\left.\left.\;\;\;\;\;\;\;\;\;\;\;\;\;\;\;\;\;\;\;\;\;\;\;\;
+\frac{16}{d-2}J_{\rm 3g}
-8\frac{3d-5}{d-2}J_{\rm 3h}\right)\right]+O(g^6)\\
-\mu^{2\epsilon}F_m
&=&
-\frac{g^4}{8}d_{\rm A}C_{\rm A}^2(I_{\rm ball}^{\rm bb}
+4J_{\rm 1a} I_{\rm 2a})+O(g^6)\\
-\mu^{2\epsilon}F_n
&=&
2d_{\rm F}\hbox{$\sum$}\!\!\!\!\!\!\int_{\{P\}}\ln P^2\\
-\mu^{2\epsilon}F_o
&=&
-\frac{Z_g^2g^2}{3}d_{\rm A}S_{\rm F}[
3(d-2)(f_1^2-2b_1 f_1)-6(d-2)J_{\rm 1a} f_1-3M^2(A_1^{\rm f}-4A_2^{\rm f})
\nonumber\\
&&
\;\;\;\;-(d-2)M^2J_{\rm 1a} f_2]+O(g^6)\\
-\mu^{2\epsilon}F_p
&=&
\frac{g^4}{4}d_{\rm A}(2S_{2\rm F}
-C_{\rm A}S_{\rm F})(d-2)[2(d-4)I_{\rm ball}^{\rm bf}
-(d-6)I_{\rm ball}^{\rm ff}+8(b_1-f_1)I_{\rm sun}^{\rm f}\nonumber\\
&&
\;\;\;\;+4J_{\rm 1a} I_{\rm sun}^{\rm f}+16J_{\rm 1a} I_{\rm 2e}]+O(g^6)\\
-\mu^{2\epsilon}F_q
&=&
g^4d_{\rm A}S_{2\rm F}\{(d-2)^2[2H_3-I_{\rm ball}^{\rm bf}-(b_1-f_1)^2 f_2]
\nonumber\\
&&
\;\;\;\;+2(d-2)J_{\rm 1a}[-I_{\rm sun}^{\rm f}-4I_{\rm 2e}
+(4-d)(b_1-f_1)f_2]\}
+O(g^6)\\
-\mu^{2\epsilon}F_r
&=&
g^4d_{\rm A}C_{\rm A}S_{\rm F}\{(d-2)[I_{\rm ball}^{\rm bf}
+2(b_1-f_1)I_{\rm sun}^{\rm f}]\nonumber\\
&&
\;\;\;\;+4J_{\rm 1a}[(d-3)I_{\rm 2c}+2I_{\rm 2d}+(d-2)I_{\rm 2e}
+(d-3)f_1b_2]\}
+O(g^6)
\end{eqnarray}
The sum of those parts in the contributions above that lead to $g^5$
(and potentially $g^5\ln g$) terms as $\epsilon\rightarrow 0$ is
\begin{eqnarray}
-\mu^{2\epsilon}F^{(5)}
&=&
-\frac{1}{12}d_{\rm A} M^2J_{\rm 1a}
\{6(Z_g^2-1)+g^2[(d^2-13d+58)C_{\rm A} b_2-4(d-2)S_{\rm F} f_2]\}\nonumber\\
&&
-2g^4(d-2)(d-4)d_{\rm A}S_{2\rm F}J_{\rm 1a}(b_1-f_1)f_2\nonumber\\
&&
+g^4d_{\rm A}C_{\rm A}^2\left\{
-\frac{d-3}{8}\frac{J_{\rm 1a}^3}{M^2}
+J_{\rm 1a}\left[-\frac{3d-5}{4(d-2)}J_{\rm 2a}+\frac{d-3}{d-2}J_{\rm 2c}
+2J_{\rm 2d}\right]
\right.\nonumber\\
&&
\;\;\;\;\;\;\;\;\;\;\;\;\;\;\;
+\left[\frac{3d-5}{16(d-2)}J_{\rm 3a}+\frac{(d-3)(3d-5)}{8(d-2)}J_{\rm 3b}
+\frac{2d-3}{2(d-2)}J_{\rm 3c}\right.\nonumber\\
&&
\;\;\;\;\;\;\;\;\;\;\;\;\;\;\;\;\;\;\;\;\;
\left.\left.-4J_{\rm 3e}+4J_{\rm 3f}+\frac{1}{d-2}J_{\rm 3g}
-\frac{3d-5}{2(d-2)}J_{\rm 3h}+J_{\rm 3i}\right]\right\}\,,
\end{eqnarray}
where $Z_g^2$ should be used up to order $g^2$. Using the identities
of Appendix \ref{basic integrals} we can simplify this expression and
get
\begin{eqnarray}
-\mu^{2\epsilon}F^{(5)}
&=&
-\frac{1}{12}d_{\rm A} M^2J_{\rm 1a}
\{6(Z_g^2-1)+g^2[(d^2-13d+58)C_{\rm A} b_2-4(d-2)S_{\rm F} f_2]\}\nonumber\\
&&
-2g^4(d-2)(d-4)d_{\rm A}S_{2\rm F}J_{\rm 1a}(b_1-f_1)f_2\nonumber\\
&&
+g^4d_{\rm A}C_{\rm A}^2\left\{
\frac{3d^2-24d+37}{8(d-7)(d-5)^2}J_{\rm 3a}
+\frac{(d-3)^3(3d-14)}{4(d-5)(d-4)(2d-9)}\right.J_{\rm 3b}\nonumber\\
&&\hspace{57pt}
\left.-\frac{(d^5-23d^4+199d^3-809d^2+1548d-1132)(d-3)}{8(d-7)(d-5)^2(d-4)^2}
\frac{J_{\rm 1a}^3}{M^2}
\right\}\,.
\end{eqnarray}
Note that for this term the cancellation of overlapping
double-frequency sum-integrals still holds outside of $d=4$.
Using the results of Appendix \ref{basic integrals} it is further
easy to see how the $1/\epsilon$ terms associated with the scales $T$
and $gT$ cancel separately so that for $\epsilon\rightarrow 0$ no
$g^5\ln g$ term arises in $F$.

\section{Basic Integrals}
\label{basic integrals}
Here we give the definitions for the integrals appearing in our
derivations.
In the next subsection, we first provide the definitions of the
sum-integrals which have been evaluated in Ref.\ \cite{ArZh} and
give the results for those that are relevant for the $g^5$ term
of the free energy density.
Then we define and give the results for five additional two-loop
sum-integrals which appear in the result of individual diagrams
but cancel each other after summing up the diagrams.
In the second subsection, the definitions of and results for the
three-dimensional integrals arising in the $g^5$ evaluation are given.
They are evaluated in Appendix \ref{3-dim evaluations}.

\subsection{Some Sum-integrals}

Here is a list of integrals evaluated in Refs.\ \cite{BrFrTa,PaCo}.
One-loop integrals are
\begin{eqnarray}
b_n&\equiv&\hbox{$\sum$}\!\!\!\!\!\!\int_P \frac{1}{P^{2n}}
=\frac{(2\pi T)^{4-2n}}{8\pi^{5/2}}
\left(\frac{\mu^2}{\pi T^2}\right)^\epsilon
\frac{\Gamma(n-\frac{3}{2}
+\epsilon)\zeta(2n-3+2\epsilon)}{\Gamma(n)}\nonumber\\
f_n&\equiv&\hbox{$\sum$}\!\!\!\!\!\!\int_{\{P\}}\frac{1}{P^{2n}}
=(2^{2n-3+2\epsilon}-1)b_n\,.
\end{eqnarray}
The relevant cases are
\begin{eqnarray}
b_1
&=&
\frac{T^2}{12}\left[1+2\epsilon\left(1+\ln\frac{\bar{\mu}}{4\pi T}
+\frac{\zeta'(-1)}{\zeta(-1)}
\right)\right]+O(\epsilon^2)\nonumber\\
b_2
&=&
\frac{1}{(4\pi)^2}\left(\frac{1}{\epsilon}+2\gamma_{\rm\scriptscriptstyle E}
+2\ln\frac{\bar{\mu}}{4\pi T}
\right)+O(\epsilon)\nonumber\\
f_1
&=&
-\frac{T^2}{24}\left[1+2\epsilon\left(1-\ln 2+\ln\frac{\bar{\mu}}{4\pi T}
+\frac{\zeta'(-1)}{\zeta(-1)}
\right)\right]+O(\epsilon^2)\nonumber\\
f_2
&=&
\frac{1}{(4\pi)^2}\left(\frac{1}{\epsilon}+2\gamma_{\rm\scriptscriptstyle E}
+4\ln 2+2\ln\frac{\bar{\mu}}{4\pi T}\right)+O(\epsilon)\,.
\end{eqnarray}

Two-loop sum-integrals are
\begin{eqnarray}
    I_{\rm sun}^{\rm b} &\equiv&
\hbox{$\sum$}\!\!\!\!\!\!\int_{PQ}{1 \over P^2 Q^2 (P + Q)^2}=0
\nonumber\\
    I_{\rm sun}^{\rm f} &\equiv&
\hbox{$\sum$}\!\!\!\!\!\!\int_{P\{Q\}} {1 \over P^2 Q^2 (P + Q)^2}=0\,.
\end{eqnarray}

Three-loop integrals are
\begin{eqnarray}
    I_{\rm ball}^{\rm bb} &\equiv&
\hbox{$\sum$}\!\!\!\!\!\!
\int_{PQK} {1 \over P^2 Q^2 K^2 (P + Q + K)^2}
\nonumber\\
    I_{\rm ball}^{\rm bf} &\equiv&
\hbox{$\sum$}\!\!\!\!\!\!
\int_{PQ\{K\}} {1 \over P^2 Q^2 K^2 (P + Q + K)^2}
\nonumber\\
    I_{\rm ball}^{\rm ff} &\equiv&
\hbox{$\sum$}\!\!\!\!\!\!\int_{\{P\}\{Q\}\{K\}}
	{1 \over P^2 Q^2 K^2 (P + Q + K)^2}
\nonumber\\
    H_3 &\equiv& \hbox{$\sum$}\!\!\!\!\!\!\int_{\{P\}QK}
	{Q \cdot K \over P^2 Q^2 K^2 (P+Q)^2 (P+K)^2} \,.
\end{eqnarray}

Now we define some integrals that were computed in Ref.\ \cite{ArZh}
but not explicitly defined:
\begin{eqnarray}
    A_1^{\rm b} &\equiv& \hbox{$\sum$}\!\!\!\!\!\!\int_{PQ}
	{ \delta_{p_0}(1 - \delta_{q_0}) \over P^2 Q^2 (P + Q)^2}
\nonumber\\
    A_2^{\rm b} &\equiv& \hbox{$\sum$}\!\!\!\!\!\!\int_{PQ} { \delta_{p_0}
	\over P^4} \left [{q_0^2 \over Q^4} -
	{q_0^2 \over Q^2 (P + Q)^2} \right ]
\nonumber\\
    A_1^{\rm f} &\equiv& \hbox{$\sum$}\!\!\!\!\!\!\int_{P\{Q\}}
	{ \delta_{p_0} \over P^2 Q^2 (P + Q)^2}
\nonumber\\
    A_2^{\rm f} &\equiv&
\hbox{$\sum$}\!\!\!\!\!\!\int_{P\{Q\}}{ \delta_{p_0}
	\over P^4} \left [{q_0^2 \over Q^4} -
	{q_0^2 \over Q^2 (P + Q)^2} \right ]
\nonumber\\
    I_{\rm qcd} &\equiv&
\hbox{$\sum$}\!\!\!\!\!\!\int_P {1 \over P^4} {\rm tr}
	\left [\Pi_{\mu\nu}(P) - \Pi_{\mu\nu}(0) \right ]^2 \,,
\end{eqnarray}
where the values for the first four integrals may be found
in the evaluations of $\delta_1,\delta_2,\delta_3$ in
Ref.\ \cite{ArZh} and $I_{\rm qcd}$ may be expressed in terms
of $I_{\rm qcd}^{\rm bb}$, $I_{\rm qcd}^{\rm bf}$,
and $I_{\rm qcd}^{\rm ff}$ there.

The following five two-loop sum-integrals appear only in individual
diagrams but not in the final result for the free energy density.
They can be evaluated using the methods introduced in \cite{ArZh}.
Here we only give the definitions and the values of these five integrals.
\begin{eqnarray}
I_{\rm 2a}&\equiv&
\hbox{$\sum$}\!\!\!\!\!\!\int_{PQ}\frac{p_0^2}{P^4Q^2(P{+}Q)^2}=0\nonumber\\
I_{\rm 2b}&\equiv&
\hbox{$\sum$}\!\!\!\!\!\!\int_{PQ}\frac{q_0^2}{P^4Q^2(P{+}Q)^2}
=\frac{T^2}{12(4\pi)^2}+O(\epsilon)\nonumber\\
I_{\rm 2c}&\equiv&
\hbox{$\sum$}\!\!\!\!\!\!\int_{P\{Q\}}\frac{p_0^2}{P^4Q^2(P{+}Q)^2}
=-\frac{T^2}{12(4\pi)^2}\left[
\frac{3}{4\epsilon}{+}\frac{3}{4}{+}\frac{3}{2}
\gamma_{\rm\scriptscriptstyle E}{+}\frac{5}{2}\ln2
{+}3\ln\frac{\bar\mu}{4\pi T}
{+}\frac{3}{2}\frac{\zeta'(-1)}{\zeta(-1)}\right]+O(\epsilon)\nonumber\\
I_{\rm 2d}&\equiv&
\hbox{$\sum$}\!\!\!\!\!\!\int_{P\{Q\}}\frac{q_0^2}{P^4Q^2(P{+}Q)^2}
=-\frac{T^2}{12(4\pi)^2}\left[
\frac{3}{8\epsilon}{+}\frac{7}{8}
{+}\frac{3}{4}\gamma_{\rm\scriptscriptstyle E}{+}\frac{5}{4}\ln2
{+}\frac{3}{2}\ln\frac{\bar\mu}{4\pi T}
{+}\frac{3}{4}\frac{\zeta'(-1)}{\zeta(-1)}\right]+O(\epsilon)\nonumber\\
I_{\rm 2e}&\equiv&
\hbox{$\sum$}\!\!\!\!\!\!\int_{\{P\}Q}\frac{p_0q_0}{P^4Q^2(P+Q)^2}
=\frac{T^2}{16(4\pi)^2}+O(\epsilon)\,.
\end{eqnarray}

\subsection{Three-dimensional Momentum Integrals}

Here is a list of our basic three-dimensional momentum integrals.
We use dimensional regularization to control both the ultraviolet
and the infrared divergences.
Therefore, ``three-dimensional momentum integrals'' really means
integrals in $3-2\epsilon$ dimensions.
The steps for computing these integrals are provided in Appendix
\ref{3-dim evaluations}.
\begin{eqnarray}
J_{\rm 1a}
&\equiv&
\hbox{$\sum$}\!\!\!\!\!\!\int_{P} {\delta_{p_0} \over P^2+M^2}\nonumber\\
&=&-\frac{TM}{4\pi}\left[1+2\left(\ln\frac{\bar{\mu}}{2M}
+1\right)\epsilon+\left(2\ln^2\frac{\bar{\mu}}{2M}
+4\ln\frac{\bar{\mu}}{2M}+4+\frac{\pi^2}{4}
\right)\epsilon^2\right]+O(\epsilon^3)\nonumber\\
    J_{\rm 2a} &\equiv&
\hbox{$\sum$}\!\!\!\!\!\!\int_{PQ} {\delta_{p_0} \delta_{q_0}
	\over (P^2 + M^2) (Q^2 + M^2) (P + Q)^2}
= {T^2 \over (4 \pi)^2} \left [{1 \over 4 \epsilon}
        + \ln {\bar\mu \over 2 M} + {1 \over 2}  \right ]+ O(\epsilon)
\nonumber\\
    J_{\rm 2b} &\equiv&
\hbox{$\sum$}\!\!\!\!\!\!\int_{PQ} {\delta_{p_0} \delta_{q_0}
        \over (P^2 + M^2) Q^2 (P + Q)^2}
= {T^2 \over (4 \pi)^2} \left [{1 \over 4 \epsilon}
        + \ln {\bar\mu \over 2M}+\ln 2 + {1 \over 2}\right ]+ O(\epsilon)
\nonumber\\
    J_{\rm 2c} &\equiv&
\hbox{$\sum$}\!\!\!\!\!\!\int_{PQ} { M^2 \delta_{p_0} \delta_{q_0}
        \over P^4 (Q^2 + M^2) [(P + Q)^2+M^2]}
= - {1 \over 8} {T^2 \over (4 \pi)^2} + O (\epsilon)
\nonumber\\
    J_{\rm 2d} &\equiv&
\hbox{$\sum$}\!\!\!\!\!\!\int_{PQ} { M^2 \delta_{p_0} \delta_{q_0}
        \over (P^2 + M^2)^2 (Q^2+M^2) (P + Q)^2}
= {1 \over 4} {T^2 \over (4 \pi)^2} + O(\epsilon)
\nonumber\\
    J_{\rm 3a} &\equiv&
\hbox{$\sum$}\!\!\!\!\!\!\int_{PQK} { \delta_{p_0} \delta_{q_0}
        \delta_{k_0} \over (P^2{+}M^2)(Q^2{+}M^2)
        (K^2{+}M^2) [(P{+}Q{+}K)^2 {+} M^2]}\nonumber\\
&=&
- {T^3 M \over (4 \pi)^3}
        \left [{1 \over \epsilon} + 6 \ln {\bar\mu \over 2 M}
	- 4 \ln 2 + 8 \right ] + O (\epsilon)
\nonumber\\
    J_{\rm 3b} &\equiv&
\hbox{$\sum$}\!\!\!\!\!\!\int_{PQK} { \delta_{p_0} \delta_{q_0}
\delta_{k_0} \over (P^2{+}M^2)(Q^2{+}M^2) K^2 (P{+}Q{+}K)^2}\nonumber\\
&=&
-\frac{T^3 M}{2(4\pi)^3}\left[\frac{1}{\epsilon}
+\left(6\ln\frac{\bar{\mu}}{2M}+8\right)
+\left(18\ln^2\frac{\bar{\mu}}{2M}+48\ln\frac{\bar{\mu}}{2M}+52
+\frac{25\pi^2}{12}\right)\epsilon\right]+O(\epsilon^2)\nonumber\\
    J_{\rm 3c} &\equiv&
\hbox{$\sum$}\!\!\!\!\!\!\int_{PQK} { M^2 \delta_{p_0} \delta_{q_0}
        \delta_{k_0} \over K^2(P^2{+}M^2)(Q^2{+}M^2)
        [(K{+}P)^2{+}M^2][(K{+}Q)^2{+}M^2] }
= {T^3 M \over (4 \pi)^3} \ln 2 + O (\epsilon)
\nonumber\\
    J_{\rm 3d} &\equiv&
\hbox{$\sum$}\!\!\!\!\!\!\int_{PQK} { M^2 \delta_{p_0} \delta_{q_0}
        \delta_{k_0} \over K^2(P^2{+}M^2)(Q^2{+}M^2) (K{+}P)^2(K{+}Q)^2 }
= {T^3 M \over (4 \pi)^3} 2 \ln 2 + O (\epsilon)
\nonumber\\
    J_{\rm 3e} &\equiv&
\hbox{$\sum$}\!\!\!\!\!\!\int_{PQK} { M^2 \delta_{p_0} \delta_{q_0}
        \delta_{k_0} \over (K^2{+}M^2)(P^2{+}M^2)(Q^2{+}M^2)
	(K{+}P)^2(K{+}Q)^2 }
= {T^3 M \over (4 \pi)^3} {\pi^2 \over 12} + O (\epsilon)
\nonumber\\
    J_{\rm 3f} &\equiv&
\hbox{$\sum$}\!\!\!\!\!\!\int_{PQK} { M^4 \delta_{p_0} \delta_{q_0}
        \delta_{k_0} \over (K^2{+}M^2)^2(P^2{+}M^2)(Q^2{+}M^2)
	(K{+}P)^2(K{+}Q)^2 }
= {T^3 M \over (4 \pi)^3}
        \left ({\pi^2 \over 24} - {1 \over 4} \right )
        + O(\epsilon)
\nonumber\\
    J_{\rm 3g} &\equiv&
\hbox{$\sum$}\!\!\!\!\!\!\int_{PQK} { M^4 \delta_{p_0} \delta_{q_0}
        \delta_{k_0} \over K^4(P^2{+}M^2)(Q^2{+}M^2)[(K{+}P)^2+M^2]
	[(K{+}Q)^2{+}M^2]} \nonumber\\
&=&
-{T^3 M \over (4 \pi)^3}
        \left(\frac{1}{24} + \frac{\ln 2}{12}\right) + O(\epsilon)
\nonumber\\
    J_{\rm 3h} &\equiv&
\hbox{$\sum$}\!\!\!\!\!\!\int_{PQK} { M^2 \delta_{p_0} \delta_{q_0}
    \delta_{k_0} \over K^2(P^2{+}M^2) [(K{+}P)^2{+}M^2] Q^2(K{+}Q)^2 }
= - {T^3 M \over (4 \pi)^3} \left [{1 \over 8 \epsilon}
	+{3 \over 4} \ln {\bar \mu \over 2 M}
	-{1 \over 4} \right ] + O(\epsilon)
\nonumber\\
    J_{\rm 3i} &\equiv&
\hbox{$\sum$}\!\!\!\!\!\!\int_{PQK} { M^4 \delta_{p_0} \delta_{q_0}
	\delta_{k_0} \over (P^2{+}M^2)[(P{+}K)^2{+}M^2](Q^2{+}M^2)
	[(Q{+}K)^2{+}M^2] K^2 (P{+}Q{+}K)^2}\nonumber\\
&=&
{T^3 M \over (4 \pi)^3} \left(\frac{1}{4} - \frac{\ln 2}{4}\right)
	+ O(\epsilon)
\nonumber\\
J_{\rm 3j}
&\equiv&
\hbox{$\sum$}\!\!\!\!\!\!\int_{PQK}{\delta_{p_0}\delta_{q_0}\delta_{k_0}
(P\cdot Q) \over (K^2{+}M^2)(P^2{+}M^2)(Q^2{+}M^2)(K{+}P)^2(K{+}Q)^2 }
=-\frac{T^3 M}{(4\pi)^3}\left(\frac{\pi^2}{12}
+\frac{1}{4}\right)+O(\epsilon)\nonumber\\
\end{eqnarray}

\section{Evaluation of Basic Integrals}
\label{3-dim evaluations}

Here we will first evaluate $J_{\rm 1a}$, $J_{\rm 3a}$ and $J_{\rm 3b}$
and then express all other three-dimensional integrals appearing in the
diagrams (a)--(r) in terms of these three.

\subsection{\protect\boldmath $J_{\rm\bf 1a}$}

$J_{\rm 1a}$ may be evaluated as
\begin{eqnarray}
    J_{\rm 1a} &=& T \mu^{2\epsilon}
\int {d^{3-2\epsilon} p \over (2\pi)^{3-2\epsilon}} {1 \over p^2 + M^2}
	= T M \left (\mu \over M \right )^{2 \epsilon}
	\int {d^{3-2\epsilon} p \over (2\pi)^{3-2\epsilon}}
\int_0^\infty ds\,e^{-s(p^2 + 1)}
\nonumber\\
	&=& T M \left (\mu \over M \right )^{2 \epsilon}
	\int_0^\infty ds\,s^{-3/2+\epsilon} e^{-s}
	= {T M \over (4 \pi)^{3/2}}
	\left (M^2 \over 4 \pi \mu^2 \right )^{-\epsilon}
	\Gamma(\textstyle-{1 \over 2} + \epsilon)
\nonumber\\
&=&
-\frac{TM}{4\pi}\left[1+2\left(\ln\frac{\bar{\mu}}{2M}
+1\right)\epsilon+\left(2\ln^2\frac{\bar{\mu}}{2M}
+4\ln\frac{\bar{\mu}}{2M}+4+\frac{\pi^2}{4}
\right)\epsilon^2\right]+O(\epsilon^3)\,.
\end{eqnarray}

\subsection{\protect\boldmath $J_{\rm\bf 3a}$ and $J_{\rm\bf 3b}$}

In dimensional regularization, we have
\begin{equation}
    J_{\rm 3a} - 2 J_{\rm 3b} =
\hbox{$\sum$}\!\!\!\!\!\!\int_K \delta_{k_0}
	\left [\hbox{$\sum$}\!\!\!\!\!\!\int_P \left (
	{\delta_{p_0} \over (P^2+M^2) [(P + K)^2 + M^2]}
	- {\delta_{p_0} \over P^2 (P + K)^2} \right ) \right ]^2\,.
\end{equation}
We are now going to evaluate this difference using a method similar
to the one applied in the appendix of the second reference of \cite{BrNi1}.

Since $J_{\rm 3a}-2J_{\rm 3b}$ is both infrared and ultraviolet finite,
we can go to three-dimensional coordinate space to get
\begin{equation}
    J_{\rm 3a} - 2 J_{\rm 3b} = T^3 \int d^3 r \left [
	{e^{-2 M r} \over (4 \pi r)^2}
	- {1 \over (4 \pi r)^2} \right ]^2 + O (\epsilon)\,,
\end{equation}
where we have used the Fourier transform
\begin{equation}
    \int {d^3 p \over (2 \pi)^3}
	{e^{i {\vec p} \cdot {\vec r}}
	\over p^2 + m^2} = {e^{-m r} \over 4 \pi r}
\label{coordinate_prop}
\end{equation}
for $m=M$ and $m=0$.
Integrating by parts and using the identity
\begin{equation}
    \int_0^\infty dr {1 \over r} (e^{-a r} - e^{-b r})
	= \ln {b \over a}
\label{logba}
\end{equation}
gives
\begin{equation}
J_{\rm 3a}-2J_{\rm 3b}={T^3 M \over (4 \pi)^3}4\ln 2+O(\epsilon)\,.
\label{J_3a_J_3b}
\end{equation}

$J_{\rm 3b}$ may be evaluated as
\begin{eqnarray}
    J_{\rm 3b} &=& T^3 M \left (\mu \over M \right )^{6 \epsilon}
	\int {d^{3-2\epsilon} p \over (2\pi)^{3-2\epsilon}}
\int {d^{3-2\epsilon} q \over (2\pi)^{3-2\epsilon}} {1 \over q^2(q+p)^2}
	\int {d^{3-2\epsilon} k \over (2\pi)^{3-2\epsilon}}
{1 \over (k^2 + 1)[(k+p)^2 + 1]}
\nonumber\\
	&=& T^3 M \left (\mu \over M \right )^{6 \epsilon}
	{\Gamma \left (\textstyle{1 \over 2}{+}\epsilon \right )^2
	\over (4 \pi)^{3{-}2\epsilon}}
	\int {d^{3-2\epsilon} p \over (2\pi)^{3-2\epsilon}}
\int_0^1 d \alpha [\alpha(1{-}\alpha) p^2]^{{-}1/2{-}\epsilon}
\!\! \int_0^1 d \beta [\beta(1{-}\beta)p^2{+}1]^{{-}1/2{-}\epsilon}
\nonumber\\
	&=& T^3 M \left (\mu \over M \right )^{6 \epsilon}
        {\Gamma \left (\textstyle{1 \over 2}{+}\epsilon \right )^2
        \over (4 \pi)^{3{-}2\epsilon}}{
	B \left (\textstyle{1 \over 2}{-}\epsilon,
	\textstyle{1 \over 2}{-}\epsilon \right )
	\over (4 \pi)^{3/2{-}\epsilon}
	\Gamma \left (\textstyle{3 \over 2}{-}\epsilon \right )}
	\int_0^1 d \beta \int_0^\infty d p^2 (p^2)^{{-}2\epsilon}
	\left[\beta(1{-}\beta)p^2{+}1\right ]^{{-}1/2{+}\epsilon}
\nonumber\\
	&=& T^3 M \left (\mu \over M \right )^{6 \epsilon}
        {\Gamma \left (\textstyle{1 \over 2}+\epsilon \right )^2
        \over (4 \pi)^{3-2\epsilon}}{
        B \left (\textstyle{1 \over 2}{-}\epsilon,
        \textstyle{1 \over 2}{-}\epsilon \right )
        \over (4 \pi)^{3/2{-}\epsilon}
        \Gamma \left (\textstyle{3 \over 2}{-}\epsilon \right )}
	B\left(1{-}2\epsilon,{-}\textstyle{1 \over 2}{+}3\epsilon\right)
	B \left (2 \epsilon , 2 \epsilon \right )
\nonumber\\
	&=& \frac{T^3 M}{(4 \pi)^{9/2}}
        \left (M^2 \over 4 \pi \mu^2 \right )^{-3 \epsilon}
	{\Gamma \left (\textstyle{1 \over 2}{+}\epsilon\right)
	\Gamma \left (\textstyle{1 \over 2}{-}\epsilon\right)^2
	\Gamma \left (-\textstyle{1 \over 2}{+}3\epsilon\right)
	\Gamma (2 \epsilon)^2 \over
	\Gamma \left (\textstyle{3\over 2}
	{-}\epsilon\right ) \Gamma(4 \epsilon)}
\nonumber\\
&=&
-\frac{T^3 M}{2(4\pi)^3}\left[\frac{1}{\epsilon}
+\left(6\ln\frac{\bar{\mu}}{2M}+8\right)
+\left(18\ln^2\frac{\bar{\mu}}{2M}+48\ln\frac{\bar{\mu}}{2M}+52
+\frac{25\pi^2}{12}\right)\epsilon\right]+O(\epsilon^2)\,.\nonumber\\
\label{J_3b_result}
\end{eqnarray}
For step two above, we have used the result
\begin{equation}
    \int {d^{3-2\epsilon} q \over (2\pi)^{3-2\epsilon}}
{1 \over (p^2 + m^2) [(p+q)^2 + m^2]}
	= {\Gamma (\textstyle{1\over 2} + \epsilon)
	\over (4 \pi)^{3 - 2 \epsilon} }
	\int_0^1 d \alpha [\alpha(1-\alpha)p^2+m^2]^{-1/2-\epsilon}\,,
\end{equation}
obtained by Feynman parametrization, for $m=0,1$.
For step three, the identity
\begin{equation}
    \int_0^1 d x x^{a-1}(1-x)^{b-1} = B (a, b)\,,
\end{equation}
where $B(a,b)$ is the Beta function
\begin{equation}
    B (a, b) = {\Gamma (a) \Gamma (b) \over \Gamma (a+b)}\,,
\end{equation}
as well as the surface area of the $d$-dimensional unit sphere,
$2\pi^{d/2}/\Gamma(d/2)$, have been used.
For step four, we used the identity
\begin{equation}
    \int_0^\infty d x {x^{a-1} \over (1 + x)^{a+b}} = B(a, b)\,.
\end{equation}
Combining the results~(\ref{J_3a_J_3b}) and (\ref{J_3b_result}) gives
the value for $J_{\rm 3a}$.

\subsection{Identities for Three-dimensional Integrals}

Here are some identities for $(3-2\epsilon)$-dimensional integrals.
It is easy to derive them and we will give a sample proof in the
following subsection.
\begin{eqnarray}
\label{J identities}
J_{\rm 2a} &=& -\frac{d-3}{2(d-4)}\frac{J_{\rm 1a}^2}{M^2}\nonumber\\
J_{\rm 2c} &=& \frac{1}{d-6}J_{\rm 2d}\nonumber\\
J_{\rm 2d} &=& -\frac{d-4}{2}J_{\rm 2a}\nonumber\\
J_{\rm 3c} &=& -\frac{1}{4(d-5)}[(3d-11)J_{\rm 3a}-4(d-3)J_{\rm 1a}
  J_{\rm 2a}]\nonumber\\
J_{\rm 3d} &=& \frac{3d-11}{d-3}J_{\rm 3b}-2J_{\rm 1a}
  J_{\rm 2b}\nonumber\\
J_{\rm 3e} &=& \frac{1}{4(d-4)}[(3d-11)J_{\rm 3b}-2(d-3)J_{\rm 1a}
  J_{\rm 2a}]\nonumber\\
J_{\rm 3f} &=& \frac{1}{2(2d-9)}[(3d-13)J_{\rm 3e}-2(d-3)J_{\rm 1a}
  J_{\rm 2d}]\nonumber\\
J_{\rm 3g} &=& \frac{1}{4(d-7)}[-(3d-13)J_{\rm 3c}+4(d-3)J_{\rm 1a}
  J_{\rm 2c}]\nonumber\\
J_{\rm 3h} &=& -\frac{3d-11}{4(2d-9)}J_{\rm 3b}\nonumber\\
J_{\rm 3i} &=& \frac{1}{4(d-5)}[(3d-13)(2J_{\rm 3e}-J_{\rm 3c})
  +4J_{\rm 3f}]\nonumber\\
J_{\rm 3j} &=& \frac{1}{4}\left[J_{\rm 3b}+J_{\rm 3d}-4J_{\rm 3e}
  -2J_{\rm 1a}(2J_{\rm 2a}-J_{\rm 2b})
           -\frac{J_{\rm 1a}^3}{M^2}\right]\,.
\end{eqnarray}
Putting all of them together lets one express all three-dimensional
integrals in terms of $J_{\rm 1a}$, $J_{\rm 2b}$, $J_{\rm 3a}$ and
$J_{\rm 3b}$:
\begin{eqnarray}
\label{J formulae}
J_{\rm 2a}  &=& -\frac{d-3}{2(d-4)}\frac{J_{\rm 1a}^2}{M^2}\nonumber\\
J_{\rm 2c}  &=& \frac{d-3}{4(d-6)}\frac{J_{\rm 1a}^2}{M^2}\nonumber\\
J_{\rm 2d}  &=& \frac{d-3}{4}\frac{J_{\rm 1a}^2}{M^2}\nonumber\\
J_{\rm 3c} &=& -\frac{1}{4(d-5)}\left[(3d-11)J_{\rm 3a}
           +\frac{2(d-3)^2}{d-4}\frac{J_{\rm 1a}^3}{M^2}\right]\nonumber\\
J_{\rm 3d} &=& \frac{3d-11}{d-3}J_{\rm 3b}-2J_{\rm 1a}J_{\rm 2b}\nonumber\\
J_{\rm 3e} &=& \frac{3d-11}{4(d-4)}J_{\rm 3b}
           +\frac{(d-3)^2}{4(d-4)^2}\frac{J_{\rm 1a}^3}{M^2}\nonumber\\
J_{\rm 3f} &=& \frac{(3d-13)(3d-11)}{8(d-4)(2d-9)}J_{\rm 3b}
           -\frac{(d-3)^2(d-5)}{8(d-4)^2}\frac{J_{\rm 1a}^3}{M^2}\nonumber\\
J_{\rm 3g} &=& \frac{(3d-13)(3d-11)}{16(d-5)(d-7)}J_{\rm 3a}
           +\frac{(d-3)^2}{8(d-7)}\left[\frac{3d-13}{(d-4)(d-5)}
           +\frac{2}{d-6}\right]\frac{J_{\rm 1a}^3}{M^2}\nonumber\\
J_{\rm 3h} &=& -\frac{3d-11}{4(2d-9)}J_{\rm 3b}\nonumber\\
J_{\rm 3i} &=& \frac{(3d-13)(3d-11)}{16(d-5)}\left[
           \frac{1}{d-5}J_{\rm 3a}+\frac{4}{2d-9}J_{\rm 3b}\right]
           +\frac{(d-3)^2(5d-23)}{8(d-4)(d-5)^2}\frac{J_{\rm 1a}^3}{M^2}
\nonumber\\
J_{\rm 3j} &=& \frac{d^2-10d+23}{4(d-4)(d-3)}J_{\rm 3b}
           -\frac{1}{4(d-4)^2}\frac{J_{\rm 1a}^3}{M^2}\,.
\end{eqnarray}
Since in the diagrams, $J_{\rm 2b}$ and $J_{\rm 3d}$ only appear through
$J_{\rm 3j}$ and therefore only in the combination
$J_{\rm 3d}+2J_{\rm 1a}J_{\rm 2b}=(3d-11)J_{\rm 3b}/(d-3)$
[see Appendix \ref{single diagrams} and Eq.\ (\ref{J identities})],
we have not bothered to write down the evaluation of $J_{\rm 2b}$, although
it is easy and can be done in general dimension.

\subsection{Proof of Identities for
\protect\boldmath $J_{\rm\bf 3c}$ and $J_{\rm\bf 3j}$}

As an example, here is the proof of the identity for $J_{\rm 3c}$ in
(\ref{J identities}). The proofs of all the other identities proceed
along the same lines with the exception of that for $J_{\rm 3j}$, which
is presented below.

Using the shorthand
\begin{equation}
\int_k\rightarrow T\mu^{2\epsilon}\int\frac{d^{d-1}k}{(2\pi)^{d-1}}
\end{equation}
we can write
\begin{eqnarray}
J_{\rm 3c}
&=&
\int_k\int_p\int_q
\frac{M^2}{k^2(p^2{+}M^2)(q^2{+}M^2)[(k+p)^2{+}M^2][(k+q)^2{+}M^2]}
\nonumber\\
&=&
-\frac{M^2}{d-1}\int_k\int_p\int_q k_i\frac{\partial}{\partial k_i}
\frac{M^2}{k^2(p^2{+}M^2)(q^2{+}M^2)[(k+p)^2{+}M^2][(k+q)^2{+}M^2]}
\nonumber\\
&=&
\frac{M^2}{d-1}\int_k\int_p\int_q\left[2+\frac{4k\cdot(k+p)}{(k+p)^2{+}M^2}
\right]
\frac{1}{k^2(p^2{+}M^2)(q^2{+}M^2)[(k+p)^2{+}M^2][(k+q)^2{+}M^2]}
\nonumber\\
&=&
\frac{1}{d-1}\left\{4J_{\rm 3c}+2M^2\int_k\int_p\int_q\left[
\frac{1}{(p^2{+}M^2)(q^2{+}M^2)[(k+p)^2{+}M^2]^2[(k+q)^2{+}M^2]}
\right.\right.\nonumber\\
&&\hspace{141pt}
\left.\left.-\frac{1}{k^2(q^2{+}M^2)[(k+p)^2{+}M^2]^2[(k+q)^2{+}M^2]}
\right]\right\}\nonumber\\
&=&
\frac{2}{d-5}\left(-\frac{1}{4}M^2\frac{\partial J_{\rm 3a}}{\partial M^2}
+M^2\frac{\partial J_{\rm 1a}}{\partial M^2}J_{\rm 2a}\right)\nonumber\\
&=&
\frac{2}{d-5}\left(-\frac{3d-11}{8}J_{\rm 3a}
+\frac{d-3}{2}J_{\rm 1a}J_{\rm 2a}\right)\,,
\end{eqnarray}
where for the second equality we have used the identity
$(\partial k_i/\partial k_i)=d-1$ and integrated by parts and where for
the last step it has been used that by dimensional considerations
$J_{\rm 3a}\propto M^{3d-11}$ and $J_{\rm 1a}\propto M^{d-3}$.

Finally, here is the proof of the identity for $J_{\rm 3j}$. Noting that
\begin{equation}
\int_p\frac{{\vec p}}{(p^2{+}M^2)(k+p)^2}
\end{equation}
is parallel to ${\vec k}$, we have
\begin{eqnarray}
J_{\rm 3j}
&=&
\int_k\int_p\int_q
\frac{p\cdot q}{(k^2{+}M^2)(p^2{+}M^2)(q^2{+}M^2)(k+p)^2(k+q)^2}\nonumber\\
&=&
\int_k\int_p\int_q
\frac{(k\cdot p)(k\cdot q)}{k^2(k^2{+}M^2)(p^2{+}M^2)(q^2{+}M^2)(k+p)^2
(k+q)^2}\nonumber\\
&=&
\frac{1}{4}\left[J_{\rm 3b}+J_{\rm 3d}-4J_{\rm 3e}
-2J_{\rm 1a}(2J_{\rm 2a}-J_{\rm 2b})
-\frac{J_{\rm 1a}^3}{M^2}\right]\,.
\end{eqnarray}

\setlength{\baselineskip}{14pt}

\end{document}